\documentclass{Interspeech}

\interspeechcameraready

\title{Streaming Sortformer: Speaker Cache-Based Online Speaker Diarization with Arrival-Time Ordering}
\author[affiliation={}]{Ivan}{Medennikov}
\author[affiliation={}]{Taejin}{Park}
\author[affiliation={}]{Weiqing}{Wang}
\author[affiliation={}]{He}{Huang}
\author[affiliation={}]{Kunal}{Dhawan}
\author[affiliation={}]{Jinhan}{Wang}
\author[affiliation={}]{Jagadeesh}{Balam}
\author[affiliation={}]{Boris}{Ginsburg}

\affiliation[nocounter]{}{NVIDIA}{USA}
\email{\{imedennikov,taejinp,weiqingw,heh,kdhawan,jinhanw,jbalam,bginsburg\}@nvidia.com}

\usepackage{microtype}
\usepackage{graphicx}
\usepackage{subfigure}
\usepackage{booktabs} 
\usepackage{mathtools}
\usepackage{amsthm}
\usepackage{comment}
\usepackage{amsmath,graphicx,amsfonts,pifont}
\usepackage{marvosym}
\usepackage{booktabs}   
\usepackage{arydshln}   
\usepackage{algorithm}
\usepackage{algorithmicx}
\usepackage{algpseudocode} 

\usepackage{amssymb}
\usepackage{pifont}
\newcommand{\cmark}{\ding{51}}%
\newcommand{\xmark}{\ding{55}}%
\usepackage{booktabs}

\usepackage{multicol}
\usepackage{microtype}      
\usepackage{graphicx}
\usepackage{multirow}
\usepackage{enumitem}
\usepackage{soul,color}
\usepackage{array}
\usepackage{footnote}
\usepackage{makecell}
\usepackage{marvosym}
\usepackage{color}
\usepackage{graphicx}
\usepackage{subfigure}
\usepackage{xcolor}
\usepackage[T1]{fontenc}
\usepackage{booktabs}
\usepackage{enumitem}

\usepackage[utf8]{inputenc}
\newcommand{\bs}[1]{\boldsymbol{-1}} 
\newcommand{\addcomment}[1]

\usepackage[english]{babel}
\usepackage{amsthm}
\usepackage{soul}
\usepackage{blindtext}
\usepackage{tablefootnote}

\begin{document}

\maketitle


\begin{abstract} 
This paper presents a streaming extension for the Sortformer speaker diarization framework, whose key property is the arrival-time ordering of output speakers.
The proposed approach employs an Arrival-Order Speaker Cache (AOSC) to store frame-level acoustic embeddings of previously observed speakers.
Unlike conventional speaker-tracing buffers, AOSC orders embeddings by speaker index corresponding to their arrival time order, and is dynamically updated by selecting frames with the highest scores based on the model's past predictions.
Notably, the number of stored embeddings per speaker is determined dynamically by the update mechanism, ensuring efficient cache utilization and precise speaker tracking.
Experiments on benchmark datasets confirm the effectiveness and flexibility of our approach, even in low-latency setups.
These results establish Streaming Sortformer as a robust solution for real-time multi-speaker tracking and a foundation for streaming multi-talker speech processing.
\end{abstract}
\section{Introduction}
As the accuracy of Automatic Speech Recognition~(ASR) systems continues to improve, the demand for robust speaker diarization frameworks has grown significantly.
This has spurred increasing interest in developing diarization systems capable of operating seamlessly in live, streaming environments.
The ability to accurately tag speakers in real-time transcriptions is critical for a wide range of applications, including live captioning, virtual meetings, and conversational analytics.
Despite its potential, streaming speaker diarization remains relatively underexplored compared to offline diarization.
Furthermore, the performance gap between offline and streaming speaker diarization is significantly wider than the gap observed between offline and online ASR systems.
This disparity underscores the strong need for further research to advance the state of the art in streaming speaker diarization.

Recently, end-to-end speaker neural diarization~(EEND) systems have gained popularity due to their improved performance and ease of use. In~\cite{fujita2019eendpfo,fujita2019eendsa}, diarization was framed as a frame-wise multi-class classification problem using permutation invariant training loss~\cite{yu2017permutation}. However, these systems were constrained by a fixed output class dimension. To address this limitation, \cite{fujita2020neural} and \cite{takashima2021end} adopted a chain-rule paradigm for sequential output, accommodating varying speaker numbers.
Horiguchi et al.~\cite{horiguchi2020end,horiguchi2022eend_eda} introduced EEND-EDA, which employs an LSTM encoder-decoder to model speaker attractors, later extending it with two-stage clustering~\cite{horiguchi2022eend_gla}. More recently, an attention-based encoder-decoder~(AED) system~\cite{chen2024attention} was proposed, incorporating multi-pass inference.

\begin{figure}[t]
  \centering
  \includegraphics[width=\linewidth]{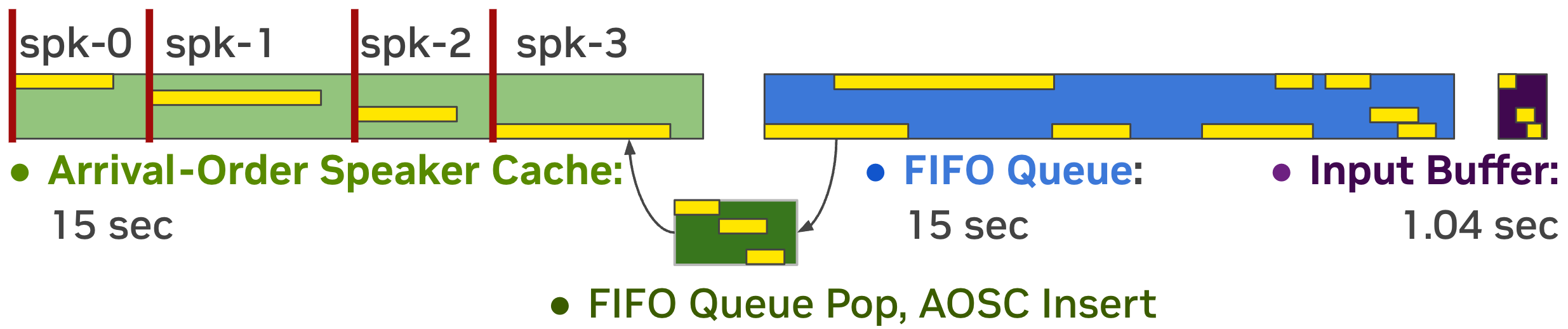}
    \vspace{-3ex}
  \caption{Speaker cache, FIFO queue and input buffer containing current chunk and right context.}
  \label{fig:fifo}
  \vspace{-5ex}
\end{figure}

For online applications, such as real-time subtitling or human-robot interaction, diarization systems must process audio streams and identify speakers in real-time.
To meet this demand, several online neural diarization systems have been developed.
Building on the offline EEND-EDA framework~\cite{horiguchi2020end,horiguchi2022eend_eda}, a block-wise version BW-EDA-EEND was introduced in~\cite{han2021bw}, which incrementally calculates speaker embeddings with a 10-second inference latency.
Subsequently, a speaker-tracing buffer~(STB)~\cite{xue2021online_flexnum,xue2021online_end} was proposed to enhance speaker consistency by storing previous frames and results, enabling low-latency inference at the cost of additional computational overhead.~In~\cite{horiguchi2022eend_gla}, unsupervised clustering was integrated into attractor-based EEND, allowing diarization for an unlimited number of speakers. Additionally, a variable chunk-size training~(VCT) mechanism was introduced to mitigate errors at the beginning of recordings.
Most recently, non-autoregressive self-attention-based attractor systems, FS-EEND~\cite{liang2024frame} and its improved version LS-EEND~\cite{liang2024ls}, have pushed the state of the art in online speaker diarization.
Moreover, TS-VAD~\cite{medennikov2020target} based streaming diarization systems~\cite{wang2023end,cheng2024sequence} are also noteworthy, showing remarkable performance in recent evaluations. 


Another promising approach recently proposed is Sortformer~\cite{park2024sortformer}, a self-attention encoder-based end-to-end speaker diarization model which differentiates itself from EEND~\cite{horiguchi2022eend_gla} based models. Sortformer employs a relatively simple architecture without attractors and is trained with Sort Loss which makes the speaker diarization model learn the arrival-time ordering for speaker predictions.
This simplifies the integration of a speaker diarization module into end-to-end multi-speaker ASR systems by eliminating the need for permutation-invariant training with token-level objectives.
However, the original Sortformer is an offline model that relies on full-length self-attention, making it unsuitable for streaming applications. Additionally, its ability to process long audio recordings is constrained by the maximum input length that the self-attention mechanism can handle, limiting its scalability for extended conversations.
\begin{figure*}[t]
  \centering
  \includegraphics[width=0.95\linewidth]{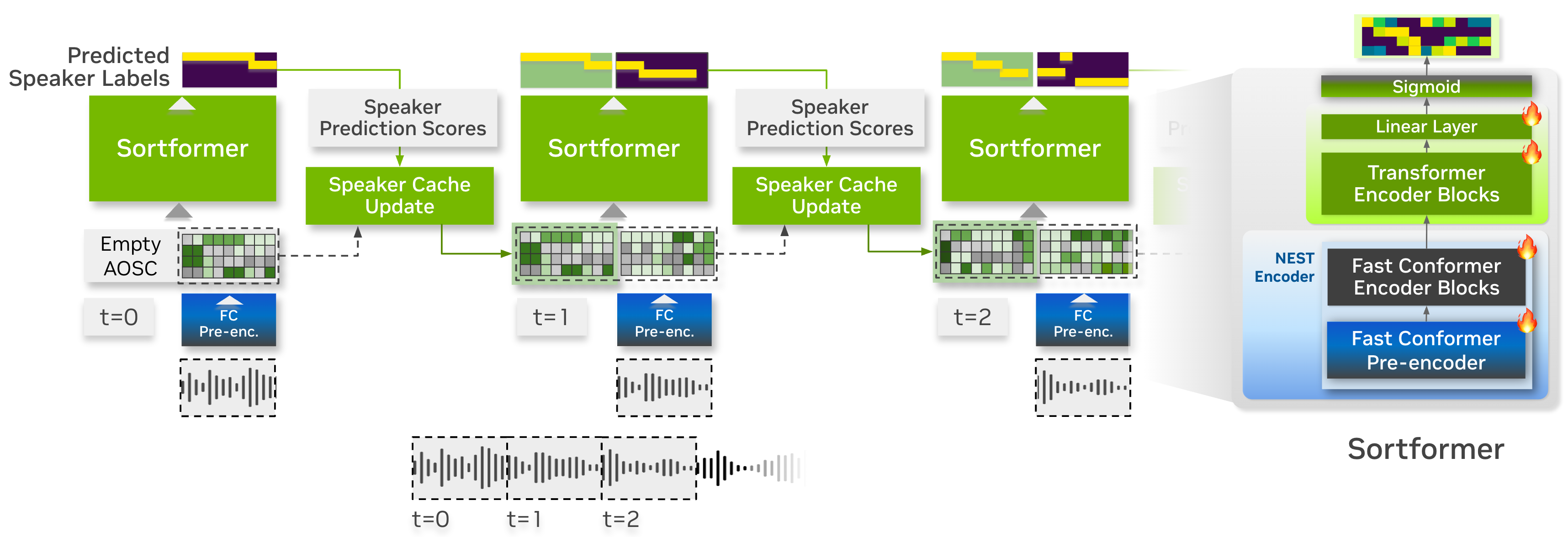}
  \vspace{-2ex}
  \caption{Streaming steps with speaker cache update. The embeddings from NEST pre-encoder are stored in the speaker cache. }
  \label{fig:streaming_steps}
  \vspace{-4ex}
\end{figure*}

In this paper, we present a simple yet effective streaming extension for the Sortformer framework. Our approach builds on the STB~\cite{xue2021online_flexnum, xue2021online_end} concept while leveraging Sortformer’s arrival-time sorting capability to resolve between-chunk permutations.
Unlike conventional STB, our memory mechanism, called Arrival-Order Speaker Cache (AOSC), accumulates acoustic embeddings in the order of speaker indices, which naturally corresponds to the arrival-time order of all previously observed speakers.
While prior works~\cite{liang2024frame, liang2024ls} also address permutation resolution using speaker appearance order, our method inherently predicts speakers in arrival-time order without relying on attractors.
As demonstrated in Figure~\ref{fig:fifo}, our proposed streaming system forms predictions at each step by concatenating the speaker cache, a number of preceding audio chunks (organized as a first-in, first-out (FIFO) queue~\cite{horiguchi2022eend_gla}), and the input buffer.
We demonstrate that this simple approach achieves superior performance without relying on self-attention attractors~\cite{liang2024frame, liang2024ls}, local or global attractors~\cite{horiguchi2022eend_gla}, or permutation resolution operations for speaker-tracing buffers~\cite{xue2021online_end, horiguchi2022eend_gla}.
Model\footnote{\url{https://huggingface.co/nvidia/diar_streaming_sortformer_4spk-v2}} and code are publicly available through the NVIDIA NeMo Framework\footnote{\url{https://github.com/NVIDIA/NeMo}}.

\section{Related Work}
\subsection{Foundation: Offline Sortformer}
The Sortformer model, as proposed in~\cite{park2024sortformer}, consists of two primary components: a self-supervised pretrained NEST encoder~\cite{huang2025nest} based on the Fast-Conformer~(FC)~\cite{rekesh2023fast} architecture, and a stack of Transformer~\cite{vaswani2017attention} encoder layers on top. The model outputs four sigmoids, allowing it to predict the activations of up to four speakers.
While the model takes Mel-spectrogram features with a 10~ms frame step as input, the convolutional pre-encode module in the NEST encoder performs 8x downsampling, resulting in an effective prediction step of 80~ms.

A key feature of Sortformer is its sorting of output speakers by their arrival time. This is achieved by the use of Sort Loss, which is a Binary Cross-Entropy computed over sorted target labels, along with the conventional Permutation-Invariant Loss.


\subsection{Chunk-Wise Processing with a Speaker-Tracing Buffer}
One common approach to streaming speaker diarization is chunk-wise processing of input audio~\cite{han2021bw,xue2021online_end,xue2021online_flexnum,horiguchi2022eend_gla}:
\begin{equation}
    \mathbf{\widehat{P}}_n = \text{EEND}(\mathbf{C}_n),
\end{equation}
where $n = 0,1,\ldots$ is chunk index, $\mathbf{C}_n = [\mathbf{X}_{nc}, \ldots, \mathbf{X}_{nc+c-1}]$ is a chunk of length $c$ from the input features $\mathbf{X} \in \mathbb{R}^{T \times D}$, and $\mathbf{\widehat{P}}_n \in \mathbb{R}^{c \times s}$ is the model's prediction for $s$ speakers in the $n$-th chunk.
The primary challenge in this approach is resolving speaker permutations across chunk-wise predictions $\mathbf{\widehat{P}}_n$ to obtain a consistently permuted sequence of predictions $\mathbf{P} = [\mathbf{P}_0, \mathbf{P}_1, \ldots]$ as the final speaker diarization output.

Previously, the speaker-tracing buffer~(STB)~\cite{xue2021online_flexnum,xue2021online_end} was successfully used to address this problem:
\begin{gather}
    \mathbf{P}_0 = \text{EEND}(\mathbf{C}_0),  \mathbf{B}_0 = \emptyset, \\
    \mathbf{P}_n^{buf}, \mathbf{B}_n = \text{STB}([\mathbf{P}_{n-1}^{buf}, \mathbf{P}_{n-1}],[\mathbf{B}_{n-1},\mathbf{C}_{n-1}]), \\
    [\mathbf{\widehat{P}}_n^{buf}, \mathbf{\widehat{P}}_n] = \text{EEND}([\mathbf{B}_n, \mathbf{C}_n]), \\
    \psi = \operatorname*{argmax}_{\phi \in \text{perm}(S)} \text{CC}(\mathbf{P}_n^{buf}, \phi(\mathbf{\widehat{P}}_n^{buf})), \\
    \mathbf{P}_n = \psi(\mathbf{\widehat{P}}_n),
\end{gather}
where $\mathbf{B}_n$ is the speaker-tracing buffer at the $n$-th step, $\mathbf{P}_n^{buf}$ is the sequence of predictions corresponding to the buffer, $\text{STB}()$ is the buffer update function, and $\psi$ is the speaker permutation that maximizes the correlation coefficient $\text{CC}$. 

\section{Proposed Method}
\subsection{Arrival-Order Speaker Cache}
While our approach also utilizes the STB concept, it introduces a specialized buffer design that aligns with Sortformer's core idea and eliminates the need for an explicit permutation resolution step.
More specifically, we propose the Arrival-Order Speaker Cache~(AOSC), which stores frame-level embeddings from the pre-encode NEST module. The key difference from STB is that these embeddings are ordered by speaker index, corresponding to the arrival-time order of speakers. Combined with Sortformer's inherent arrival-ordering mechanism, this allows for automatic resolution of between-chunk permutations:
\begin{gather}
    \mathbf{P}_0 = \text{Sortformer}(\mathbf{C}_0),  \mathbf{B}_0 = \emptyset, \\
    \mathbf{P}_n^{buf}, \mathbf{B}_n = \text{AOSC}([\mathbf{P}_{n-1}^{buf}, \mathbf{P}_{n-1}],[\mathbf{B}_{n-1},\mathbf{C}_{n-1}]), \\
    [\_, \mathbf{P}_n] = \text{Sortformer}([\mathbf{B}_n, \mathbf{C}_n]),
\end{gather}
where $\text{AOSC}()$ is the speaker cache update function, and $\_$ represents the predictions for the speaker cache at the $n$-th step, which we discard. Figure~\ref{fig:streaming_steps} illustrates the streaming processing steps in the proposed system.
   

\subsection{Speaker Cache Update Mechanism}
\label{ssec:sc_update}
The AOSC operates as a no-op function when the length of the input sequence is less than the maximum speaker cache length $M$.
Otherwise, the input must be compressed into an $M$-length sequence.
Compression is performed by retaining the embeddings for frames with the highest scores, based on the model's predictions for each frame.
Below is a step-by-step description of the update mechanism used in AOSC.
\begin{enumerate}[leftmargin=0.7cm]
\item \textbf{Compute speaker scores} $S$ for each frame:
\begin{align}
\vspace{-3ex}
    S_i &= \log P_i + \sum_{j \neq i} \log(1 - P_j),
\end{align}
where $i$ is the speaker index, and $P$ represents the model's prediction for the frame. 
\item \textbf{Detect silence frames} where the model assigns low probability to all speakers, then compute the \textbf{average silence embedding} over these frames.
\item \textbf{Disable non-speech scores}: if $P_i < 0.5$, set $S_i =  -\infty$.
\item \textbf{Prioritize recent frames}: For frames corresponding to newly added embeddings, increase their scores by $\delta > 0$ to favor keeping recent speaker data in the speaker cache.
\item \textbf{Ensure speaker representation}: For each speaker, increase $K$ highest scores by $\Delta > 0$, ensuring that each speaker is represented in the speaker cache.
\item \textbf{Append $A$ scores of $+\infty$ for each speaker}, corresponding to the average silence embedding.
\item Concatenate scores for all speakers, then \textbf{select the $M$ highest-scoring frames} and return the corresponding embeddings, while preserving their order. For frames corresponding to $+\infty$ or $-\infty$ scores, use the average silence embedding instead.
\end{enumerate}

The resulting sequence preserves embeddings for all present speakers, with these embeddings ordered by speaker index.
Additionally, $A$ silence embeddings are appended after each speaker's embeddings to facilitate speaker transition detection. These silence embeddings are computed as the average of frames where Sortformer assigns a low probability to all speakers.
It is important to note that the number of embeddings per speaker is determined dynamically based on their respective scores. However, a minimum of $K$ frames per speaker is enforced, if the corresponding scores are greater than $-\infty$.
This design ensures robust and flexible tracking of speaker profiles stored in the speaker cache, allowing the system to focus on the most relevant speech samples for each speaker.

\subsection{Streaming Inference with a FIFO Queue}
Short-chunk processing generally degrades accuracy due to limited context.
To mitigate this issue, we integrate a first-in, first-out (FIFO) queue~\cite{horiguchi2022eend_gla} alongside AOSC.
This approach not only improves context utilization but also allows AOSC updates to be performed with a larger update period, instead of after each individual chunk, improving both robustness and efficiency.
Figure~\ref{fig:fifo} illustrates the placement of the speaker cache, FIFO queue, and input buffer, which contains both the current chunk and future context. When frames stored in the FIFO queue are pushed out, they are processed by the speaker cache update mechanism.

\begin{table*}[ht!]
\centering
\caption{Diarization error rate (DER) for speaker diarization. All evaluations include overlapping speech. Collar tolerance is \SI{0}{\second} for DIHARD III Eval, and \SI{0.25}{\second} for CALLHOME-part2 and CH109. Bold numbers indicate the lowest DER among systems with \SI{1}{\second} latency.}
\vspace{-2ex}
\label{table:sd}
\setlength{\tabcolsep}{3pt} 
\begin{tabular}{rcc|ccc|cccccc|c}
\Xhline{3\arrayrulewidth}
\toprule
Diarization & Latency & Post & \multicolumn{3}{c|}{\textbf{DIHARD III Eval}} & \multicolumn{6}{c|}{\textbf{CALLHOME-\texttt{part2}}} & \textbf{CH109} \\
Systems & (sec)  & Processing &  $\leq$4 spk & $\geq$5 spk & all & 2 spk & 3 spk & 4 spk & 5 spk & 6 spk & all & 2 spk \\
\Xhline{2\arrayrulewidth}
\midrule
BW-EDA-EEND~\cite{han2021bw}                                      & 10 & -      & -     & -    & -     & 11.82 & 18.30 & 25.93 & -   & -    & -    & -    \\
EEND-EDA + FW-STB~\cite{xue2021online_end, horiguchi2022eend_gla} & 1  & -      & 19.00 & 50.21 & 25.09 & 9.08 & 13.33 & 19.36 & 30.09 & 37.21 & 14.93 & - \\
EEND-GLA-Large + BW-STB~\cite{horiguchi2022eend_gla}              & 1  & -      & 14.81 & 45.17 & 20.73 & 9.20 & 12.42 & 18.21 & 29.54 & 35.03 & 14.29 & - \\
FS-EEND+VCT~\cite{liang2024frame}                                 & 1  & -      & -     & -     & -     & 9.40 & 14.00 & 20.90 & -     & -     & -     & -  \\
LS-EEND~\cite{liang2024ls}                                        & 1  & -      & 13.96 & 42.98 & 19.61 & 7.03 & 11.59 & 15.30 & 24.63 & 27.89 & 12.11 & - \\
\midrule
\multirow{2}{*}{Offline Sortformer}            & \multirow{2}{*}{$\infty$}    & \xmark & 15.47 & 47.73 & 21.71 & 6.70  & 10.36  & 15.84 & 27.20 & 32.37 & 11.97 & 5.37 \\ 
                                                                          &   & \cmark & 14.17 & 51.51 & 21.39 & 5.82  & 9.19  & 14.25 & 31.75 & 35.38 & 11.26 & 4.86 \\
\midrule
\multirow{2}{*}{Offline Sortformer-AOSC}       & 10                           & \xmark & 21.59 & 53.18 & 27.58 & 12.80  & 16.21  & 23.19 & 34.59 & 39.64 & 18.54 & 12.82 \\ 
                                               & 1.04                         & \xmark & 22.97 & 53.11 & 29.46 & 17.44  & 20.51  & 26.96 & 36.44 & 43.77 & 22.58 & 17.90 \\
\midrule
\multirow{6}{*}{Streaming Sortformer-AOSC}     & \multirow{2}{*}{10}          & \xmark & 14.79 & 41.06 & 19.88 & 6.80  & 11.27  & 12.21 & 21.12 & 27.84 & 11.10 & 5.27 \\ 
                                               &                              & \cmark & 13.67 & 41.45 & 19.02 & 6.06  & 10.01  & 11.22 & 20.34 & 26.97 & 10.09 & 4.82 \\ \cmidrule(lr){2-13}
    & \multirow{2}{*}{1.04}                                                   & \xmark & 14.57 & \textbf{42.12} & 19.89 & 7.35  & 11.57 & 13.83 & 25.81 & 29.06 & 12.00 & 5.59 \\
                                               &                              & \cmark & \textbf{13.32} & 42.61 & \textbf{18.97} & \textbf{6.43}  & \textbf{10.26} & \textbf{12.40} & \textbf{24.41} & \textbf{27.78} & \textbf{10.79} & \textbf{5.09}  \\ 
                                               \cmidrule(lr){2-13}
    & \multirow{2}{*}{0.32}                                                   & \xmark & 14.63 & 43.76 & 20.25 & 8.60  & 13.23 & 16.08 & 28.10 & 30.63 & 13.66 & 6.60 \\
                                               &                              & \cmark & 13.43 & 43.98 & 19.32 & 6.86 & 10.84 & 13.64 & 25.78 & 28.58 & 11.50 & 5.41  \\ 

\bottomrule
\Xhline{3\arrayrulewidth}
\vspace{-5ex}
\end{tabular}
\end{table*}

\section{Experimental Results}
\label{sec:experimental_results}
\subsection{Datasets}
We adopted the training dataset from~\cite{park2024sortformer}, which consists of 5150 hours of simulated mixtures and 2030 hours of real multi-talker speech. The real speech data includes Fisher English Training Speech Part 1 and 2~\cite{cieri2004fisher}, the AMI Corpus Individual Headset Mix~(IHM)~\cite{ami} (train and dev splits from~\cite{landini2022bayesian}), DIHARD III Dev~\cite{ryant2020third}, VoxConverse-v0.3~\cite{voxconv}, ICSI~\cite{icsi}, AISHELL-4~\cite{fu2021aishell}, and NIST SRE 2000 CALLHOME \texttt{Part1}\footnote{We use two-fold splits from the Kaldi \texttt{callhome\_diarization} recipe~\cite{callhome2}, where \texttt{Part1} is used for training and fine-tuning, and \texttt{Part2} is reserved for evaluation, consistent with the methodology used in other studies we compare against.}~\cite{callhome}. 
Additionally, we included AMI Lapel Mix and Single Distant Microphone~(SDM) recordings, AliMeeting recordings from both near and far microphones~\cite{alimeeting}, and the DiPCo~\cite{segbroeck2020dipco} dataset with forced alignment-based RTTMs from~\cite{mitrofanov2024stcon}.
We cut all datasets into 90-second segments with up to four speakers, applying an 8-second shift between consecutive segments.
Also, for the AliMeeting~\cite{alimeeting} dataset, we used an offline Sortformer model to filter out segments with a high insertion rate, addressing unannotated parts of the recordings.

For evaluation, we assessed the models on DIHARD III Eval~\cite{ryant2020third}, CALLHOME \texttt{Part2}~\cite{callhome}, and CH109\footnote{While some overlap exists between CH109 recordings and CALLHOME \texttt{part1} used for training, it is very minor and should not affect the results significantly.}, a two-speaker subset of 109 sessions from the Callhome American English Speech~(CHAES) dataset~\cite{canavan1997CALLHOME}.

\subsection{Training Setup} 
For baseline offline Sortformer training, we follow the configuration described in~\cite{park2024sortformer} with minor modifications.
First, instead of the 115M-parameter NEST encoder~\cite{huang2025nest} trained on 80-dimensional Mel-spectrogram features and English speech, we use an advanced 109M-parameter version trained on multilingual data with 128-dimensional Mel-spectrograms.
Second, we remove global feature normalization, as it is unsuitable for streaming. As in~\cite{park2024sortformer}, temporal resolution of Sortformer output is 80 ms, and the maximum number of speakers is four. The total number of parameters is 117M.

The offline Sortformer checkpoint is then fine-tuned with the AOSC mechanism.
During training, we process 90-second samples in sequential 15-second windows, updating the speaker cache at each step. Speaker cache size used for training is 15 seconds (188 frames). For speaker cache update mechanism (subsection~\ref{ssec:sc_update}), parameters used are $A = 3$ frames, $\delta=0.05$. Step 5 was applied twice: strong boosting of $K=33$ frames per speaker by $\Delta=-2 \log{0.5}$, and weak boosting of $K=66$ frames per speaker by $\Delta=-\log{0.5}$. 

Notably, we do not apply common data augmentation techniques such as SpecAugment~\cite{park2019specaugment} or RIR+Noise augmentation~\cite{ko2017study}.
Instead, we introduce a random permutation of all speakers in the speaker cache at each streaming step.
Additionally, to make the model rely less on future context, we limit the right context of self-attention to 7 frames~(560~ms) with a 50\% probability for each training batch.
All model training runs are conducted with a batch size of 4 on 64×NVIDIA Tesla V100 GPUs.


\subsection{Evaluation and Comparative Analysis}
Table~\ref{table:sd} presents the evaluation results of our proposed system across several popular speaker diarization benchmark datasets.
Note that, while Sortformer is primarily designed and optimized for scenarios with up to 4 speakers, we also evaluate its performance on benchmarks with 5+ speakers to understand the system's behavior beyond its primary design scope.

We compare three variants:
\begin{enumerate}
\item \textbf{Offline Sortformer}: An offline system where the entire input audio is processed at once.
\item \textbf{Offline Sortformer-AOSC}: A system that employs AOSC only during inference, without fine-tuning.
\item \textbf{Streaming Sortformer-AOSC}: A fully streaming system fine-tuned with the AOSC module to optimize performance.
\end{enumerate}
The setup details for each latency configuration are provided in Table~\ref{table:latency_setups}.
It should be noted that the declared latency values refer to the input buffer delay and do not include algorithmic latency from computation. To quantify this algorithmic latency, we report the Real-Time Factor~(RTF), defined as the time taken to process a recording divided by its length. These RTF measurements were performed with a batch size of 1 on an NVIDIA RTX 6000 Ada Generation GPU.


\begin{table}[t]
\centering
\caption{Latency setups with corresponding parameters.}
\label{table:latency_setups}
\vspace{-2ex}
\setlength{\tabcolsep}{3pt} 
\begin{tabular}{c|ccccc|c}
\Xhline{2\arrayrulewidth}
\toprule
\multirow{2}{*}{\textbf{Latency}} & Chunk  & Right & FIFO  & Update & Speaker & \multirow{2}{*}{RTF} \\ 
    & Size & Context & Queue  & Period  & Cache &  \\ 
\big[sec\big] & \multicolumn{5}{c|}{\big[frame count\big]} & \\ \midrule
10.0 & 124  & 1  & 124 & 124  & 188 & 0.005                    \\
1.04 & 6    & 7  & 188   & 144 & 188 & 0.093                   \\
0.32 & 3    & 1  & 188   & 144 & 188 & 0.180                   \\ 
\bottomrule
\Xhline{2\arrayrulewidth}
\end{tabular}%
\vspace{-4ex}
\end{table}


It is important to emphasize that, unlike most studies that fine-tune their models separately for each evaluation dataset, our DERs reported in Table~\ref{table:sd} are achieved with a single model, without any dataset-specific fine-tuning across all evaluation sets.
However, we note that each dataset has a different DER evaluation setup, particularly with respect to collar length. To address errors arising from variations in collar length and annotation styles, we apply timestamp post-processing consisting of six operations: onset thresholding, offset thresholding, onset padding, offset padding, and the removal of short silences or speech segments below a specified threshold.
Two distinct sets of post-processing parameters were tuned on recordings with a maximum of 4 speakers: the first on the DIHARD III Dev split for DIHARD III Eval, and the second on the CALLHOME \texttt{Part1} split for CALLHOME \texttt{Part2} and CH109.



The evaluation reveals several important insights.
First, AOSC works with the offline Sortformer even without fine-tuning. However, DER significantly increases in this mode, highlighting that fine-tuning with AOSC is essential for optimal results.

Notably, streaming Sortformer demonstrates solid performance on 5+ speaker subsets, while having a constraint of 4 speakers by design. This suggests that the system is able to accurately and robustly track the four most dominant speakers in a conversation, thereby effectively managing the diarization task in these more complex scenarios.

Another notable observation is that, for DIHARD III and CALLHOME 4+ speakers, the streaming Sortformer outperforms the offline model. This likely stems from the offline Sortformer’s underperformance on long recordings due to a mismatch with 90-second training samples. In contrast, streaming Sortformer avoids this issue with its fixed inference window.

Finally, in comparison with previously published streaming systems, our system achieves state-of-the-art results.
Furthermore, although performance expectedly decreases as latency is reduced, the degradation is not severe. Even with a very low latency of 0.32 seconds, the model continues to deliver highly competitive performance. This underscores the robustness and flexibility of the proposed system.

\section{Conclusion}
\label{sec:conclusion}
\noindent
In this paper, we propose a streaming version of the Sortformer diarization model, incorporating a novel speaker cache mechanism called AOSC.
Our speaker cache management technique dynamically adjusts the cache size for each speaker, focusing on speech frames that are most valuable for caching.
The proposed streaming diarization framework achieves state-of-the-art performance on datasets such as DIHARD III and CALLHOME for scenarios with up to four speakers.
For future work, we plan to extend the system to handle up to eight speakers, broadening its applicability.
Additionally, we aim to integrate the streaming Sortformer diarizer into various multi-speaker speech processing tasks, including ASR, speech translation, and summarization, enabling real-world applications such as broadcasting and meeting transcription. 
We hope our research will inspire further advancements in multi-speaker ASR, speaker diarization, and other related speech technologies.
\newpage
\clearpage
\bibliographystyle{IEEEtran}
\bibliography{mybib,reference}

\end{document}